\documentclass[lineno]{jfm_arxiv}
\usepackage{tikz,float}
\usepackage{graphicx}
\usepackage{newtxtext}
\usepackage{newtxmath}
\usepackage{natbib}
\usepackage{hyperref}
\hypersetup{
    colorlinks = true,
    urlcolor   = blue,
    citecolor  = black,
}

\newcommand{\RomanNumeralCaps}[1]
\linenumbers
\newcommand{\rem}[1]{}

\usepackage{mathrsfs}
\newcommand{\bom}{\mbox{\boldmath$\omega$}}
\newcommand{\bxi}{\mbox{\boldmath$\xi$}}

\newcommand{\bu}{\mbox{\boldmath$u$}}

\newcommand{\br}{\mbox{\boldmath$r$}}
\newcommand{\bx}{\mbox{\boldmath$x$}} 
\newcommand{\bdf}{\mbox{\boldmath$f$}}

\newcommand{\I}{\int_{V}}

\newcommand{\bel}{\begin{equation}\label}
\newcommand{\ee}{\end{equation}}
\newcommand{\beq}{\begin{eqnarray}\label} 
\newcommand{\eeq}{\end{eqnarray}} 
\newcommand{\bc}{\begin{center}} 
\newcommand{\ec}{\end{center}} 
\newcommand{\ben}{\begin{enumerate}}
\newcommand{\een}{\end{enumerate}}
\newcommand{\bit}{\begin{itemize}}
\newcommand{\eit}{\end{itemize}}
\newcommand\sthird{\ensuremath{{\scriptstyle\frac{1}{3}}}}

\newcommand\twothirds{\ensuremath{{\scriptstyle\frac{2}{3}}}}
\newcommand\fourfifths{\ensuremath{{\scriptstyle\frac{4}{5}}}}
\newcommand{\Dim}{\mathfrak{D}_{m}}

\newcommand{\Fm}{\mathbb{F}_{m}}
\newcommand{\Sp}{S_{p}}

\title{\large Is it true that no mathematical relation exists between the Navier-Stokes equations and the multifractal model?}
\author{John D. Gibbon\aff{1}\corresp{\email{j.d.gibbon@ic.ac.uk}} \and Dario Vincenzi\aff{2}}
\affiliation{\aff{1}Department of Mathematics, Imperial College London, London SW7 2AZ, UK
\aff{2}Universit\'e C\^ote d'Azur, CNRS, LJAD, 06100 Nice, France}

\begin{document}
\maketitle
\begin{abstract}
Contrary to accepted turbulence folklore, which holds that no mathematical relation exists between the Navier-Stokes equations (NSEs) and the multifractal model (MFM) of Parisi and Frisch, we develop a theory that reconciles the MFM with Leray's weak solutions of Navier-Stokes analysis. From a combination of Euler invariant scaling and the NSEs set in a  three-dimensional box of side $L$, we also derive the the Paladin–Vulpiani scale $\eta_{h,pav}$, which is related to the Reynolds number $Re$ by $L\eta_{h,pav}^{-1} = Re^{1/(1+h)}$, and which acts as a mediator between the two theories. This is achieved by considering $L^{2m}$-norms of the velocity gradient to find a correspondence between $m$ and the local scaling exponent $h$ in the multifractal model. The parameter $m$ acts as if it were the sliding focus control on a telescope which allows us to zoom in and out on different structures. The range $1 \leq m \leq \infty$ is equivalent to $-\twothirds \leq h_{min} \leq \sthird$, which lies precisely in the region where \citet{Bandak2022,Bandak2024} have suggested that thermal noise makes the NSEs inadequate and generates spontaneous stochasticity. The implications of this are discussed. 
 \end{abstract}
 
\par\vspace{0mm}

%%%%%%%%%%%%%

\section{Introduction}\label{intro}

%%%%%%%%%%%%%%%%%%%%%%%%
\subsection{Historical perspective}

The language of fractal physics has been particularly useful in understanding high Reynolds number flow-visualizations that show the simultaneous existence and dynamic evolution of vortical structures ranging from fully three-dimensional large vortices, down to quasi-one-dimensional filaments, and even broken structures whose dimensions are less than unity. Many historical references can be found in \citet{PKY2020,SY2021} and \citet{Jim2025}\,; see also \citet{Sreeni1991,JCV2015,IGK2009,PPR2009,PKY2018,BD2019,BPBY2019,BBP2020,EIH2020,EIH2023,BT2023,McKeown2023,SSR2024,Brewer2025} and \citet{Kerr2025}. In their recent report on an extensive series of $32,768^{3}$ computations, \citet{PKY2025} have pointed out that 
\begin{quote}
\textit{\small Three-dimensional turbulence governed by the Navier–Stokes equations continues to be a grand challenge in computational science, even as leadership-class computing power has recently advanced past exascale.}
\end{quote}
\par\smallskip\noindent
Theoretical approaches involving fractal geometry and dynamical systems began to intersect in the 1970s and 1980s when conventional descriptions of fluid turbulence as merely quasiperiodic motion \citep{LL1959} began to be questioned\,: see the history described by \citet{Ruelle1995}. In their influential paper, \citet{RT1971} suggested that as the Reynolds number increased, fluid motion would pass through a generic sequence of bifurcations, which included both periodic and quasi-periodic motion, but whose end-state was chaotic motion on a strange attractor. This end-state was interpreted as `turbulence', although it is now recognized that this specific sequence is only applicable to tightly confined low-dimensional dynamical systems\,: see the review by \citet{JPE1981} and the paper by \citet{Lib1982}. The applicability of these ideas to the Navier-Stokes equations (NSEs) lay open to question because the theory expounded by \citet{Ruelle1978a,Ruelle1978b} was valid for \textit{finite} dimensional dissipative dynamical systems. The subsequent proof by \citet{FT1979} that the $2D$ Navier–Stokes equations possess a finite dimensional global attractor $\mathcal{A}$ made the connection secure in this case. Indeed, the further work by \citet{Ruelle1982,CF1985}  and \citet{Constantin1987} on Lyapunov exponents of $\mathcal{A}$ laid the basis for the sharp (to within logarithms) estimate of the Lyapunov dimension of the $2D$ Navier-Stokes global attractor \citep{CFT1988}. It must be stressed however, that this process cannot be followed through for the $3D$ NSEs. The Millenium problem of the global regularity of solutions in three dimensions remains open and therefore the existence of a finite dimensional global attractor remains open. What is known rigorously is the existence of Leray-Hopf weak solutions \citep{Leray1934,Hopf1951}, and it is the correspondence between these and the multifractal model which is the topic of this paper. 
\par\smallskip%\noindent
The wide range of scales observed in turbulent flows has traditionally been interpreted in terms of a Richardson cascade which involves the transfer of energy from large to smaller scales of motion\,: see \citet{MY1975} and also the review by \citet{Jim2000} who have discussed the effect of intermittency on cascades. \citet{Mandel1977} popularized the idea that this wide range of scales indicates an underlying fractal geometric structure, but it also became increasingly clear that the idea of a monofractal system with just a single dimension was not enough to provide an adequate description. \citet{PF1985} then interpreted these cascade processes in multifractal terms where there is a continuous spectrum of dimensions. A parameter designated as $h$, which appears naturally in the invariant scaling of the underlying Euler equations (see \eqref{rescal} below), was used as a local scaling exponent, with each value of $h$ belonging to a given fractal set of dimension $D(h)$. This has become known as the multi-fractal model (MFM) of turbulence\,: expositions can be found in \citet{MS1991,Frisch1995,BJPV1998,BMV2008,BB2009} and \citet{Frisch2016}, together with two series of lecture notes respectively by \citet{Eyink2008} and \citet{BT2023}. In fact, multifractal scaling has a broad application in the physical sciences \citep{SM1988}. While hydrodynamic turbulence is perhaps its most prominent example, other turbulent systems must be included, such as passive scalar fields \citep*{PMS1988} and bacterial suspensions \citep{LL2012}.

%%%%%%%%%%%%%%%%%%%%%%%%%%%
\subsection{A reconciliation between the MFM and the NSEs} 

It has become part of the folklore of turbulence that no mathematical relation exists between the MFM, which is applicable to statistically steady, homogeneous, isotropic turbulence (HIT), and the forced incompressible $3D$ NSEs 
\bel{nse1}
\left(\partial_{t} + \bu\cdot\nabla\right) \bu + \nabla p = \nu\Delta\bu + \bdf(\bx)\qquad \mbox{div}\,\bu = 0\,,
\ee
which are evolution equations requiring specific initial conditions. In (\ref{nse1}) $\bu$, $p$ and $\nu$ are respectively the velocity field, pressure and viscosity of the fluid, while $\boldmath{f} (\bx)$ is the body forcing. Building on piecemeal attempts in this direction \citep{DG2022,JDG2023}, this paper will demonstrate that a reconciliation between the two theories is possible. There are two ways of approaching this reconciliation, both of which involve a result of \citet{PV1987a,PV1987b} who used the MFM formalism to introduce a length scale into Navier-Stokes turbulence which we call the PaV-scale $\eta_{h,pav}$ (to avoid confusion with potential vorticity). In terms of the Reynolds number $Re$ the inverse PaV-scale is
\bel{PVscale}
\ell\eta_{h,pav}^{-1} = \mathit{Re}^{\frac{1}{1+h}}\,,
\ee
where the definition of the Reynolds number in \citet{PV1987a,PV1987b} was originally based on the energy dissipation rate and was derived using phenomenological arguments rather than directly from the NSEs.  Moreover, the velocity field was assumed to be statistically stationary.  To establish a mathematical relation between the MFM and the NSEs, it is critical that we use a definition of the PaV-scale that is not only based on the NSEs but which is also expressed in terms of a NS-based Reynolds number. Let us consider the invariant scaling of the Euler equations. Let $\eta$ be an inner length scale $0 < \eta \leq \ell$ and $\ell$ an outer scale which, in the periodic case, would be the box size $\ell = L$.  Consider the incompressible Euler equations in \textit{dimensional} variables $\bu(\bx,\,t)$. If we introduce a typical velocity $U_{0}$, a frequency $\varpi_{0} = U_{0}\ell^{-1}$ and the dimensionless parameter $\lambda = \ell\eta^{-1} \geq 1$, then the \textit{dimensionless} set of primed variables $\bu'(\bx',\,t')$ defined by ($h+1 > 0$)
\bel{rescal}
\bx' = \lambda\ell^{-1}\bx\,;\qquad t' = \lambda^{1-h}\varpi_{0}t\,; \qquad\bu = U_{0}\lambda^{-h}\bu'
\ee
transforms the Euler equations in \textit{dimensional} variables into the \textit{dimensionless} Euler equations in primed variables. While the Euler equations are invariant under the transformation in (\ref{rescal}) for every value of $\lambda \geq 1$, the NSEs are not. Instead, direct application of (\ref{rescal}) to the NSEs (\ref{nse1}) shows that $\eta=\eta_{h,pav}$ is that particular value of $\eta$ for which the NSEs in \textit{dimensional} variables $\bu(\bx,\,t)$ and with corresponding Reynolds number $Re=\ell U_{0}\nu^{-1}$, are transformed into the NSEs in \textit{dimensionless} variables $\bu'(\bx',\,t')$ with Reynolds number equal to unity
\bel{nseprim}
\left(\partial_{t'} + \bu'\cdot\nabla'\right) \bu' + \nabla'\!p' = \Delta'\bu' + \bdf'(\bx')\qquad \mbox{div}'\bu' = 0\,.
\ee
Thus $\eta=\eta_{h,pav}$ occurs exactly where the inertial and dissipative terms balance. This interpretation of $\eta_{h,pav}$ has been used, in particular, to obtain `fusion rules' for the statistical correlations between the  inertial-range increments and the derivatives of the velocity \citep{Benzi1998,Friedrich2018}. Depending on the allowed range of $h$ we see that the invariant scaling transformation of the Euler equations lies behind the onset of NS-turbulence as a multi-scale phenomenon. Kolmogorov's theory \citep{Frisch1995}, known as K41, restricts this to a single scale by insisting on $h = \sthird$ but, as we see below, the MFM allows a wider spread of $h$. Moreover, we shall also see that the PaV-scale is much more than just a bridge between the Euler and the NSEs because it also stitches together the MFM and Leray's weak solution formalism of the NSEs \citep{Leray1934,RRS2016} which is manifest as sequences of bounded time averages \citep{JDG2019,JDG2020}. 
\par\smallskip%\noindent
Our first approach, expounded in \S\ref{sect4:DG}, is to adapt the method of \citet{DG2022} for evaluating the higher norms of the velocity gradient by using the Euler scaling invariance in the probabilistic sense of the MFM at the wave-number $k_{h,pav} = \eta_{h,pav}^{-1}$.  Selecting only this wavenumber in the spectrum with a variation in $h$, and then averaging over $h$, circumvents the difficulty with the Kolmogorov picture where the spectrum is divided into the inertial range and the dissipation range, a division which conventional Sobolev methods of NS-analysis do not recognize. Comparison with time-averaged results from the NSEs in \S\ref{sect5:tele} shows that the results are consistent with the four-fifths law\,; namely that the multifractal spectrum (co-dimension) obeys the inequality $C(h) \geq 1-3h$, with $h$ lying in the range $-\twothirds \leq h \leq \sthird$. This corresponds to the dissipation range. The typical velocity field $U_{0}$ in (\ref{rescal}) is chosen as the root-mean-square velocity field of the NSEs, as in (\ref{nse2}), and the outer scale $\ell$ is chosen as the box-scale $L$ of the NSEs. The outer scale $\ell$ could equivalently be taken as the integral scale of the velocity or the characteristic length scale of the forcing. Such choices would yield an additional prefactor $(L/\ell)^3$ in the Navier--Stokes estimates provided in \S~\ref{sect3:NSEs}, but would not change the way the estimates scale with $\Rey$, which is the key element in our analysis.
\par\smallskip%\noindent
The second idea also involves the PaV-scale but in the reverse order. In \S\ref{sect5:tele} the parameter $m$ in the $L^{2m}$-norms of the velocity gradient is treated as if it were the sliding focus control of a telescope through which one can zoom in and out on intermittent events. This allows us to mimic weak solution time averages as if they exist on a multifractal set $\Fm$ whose range of dimensions is $\Dim=3/m$. Under the assumption that the NS-flow has settled into a fully developed state, $\Dim$ can then be compared with $D(h)$. This gives a simple relation between the scaling exponent $h$ and $m$. The range $1 \leq m \leq \infty$ corresponds to $-\twothirds \leq h_{min} \leq \sthird$, which gives a cut-off deep in the dissipation range. The PaV-scale $\eta_{h,pav}$ then re-emerges as the best estimate over all $m$ for the natural length scale of the problem. 

%%%%%%%%%%%%%%%%%%%%%%%%
\subsection{Connection with two open questions}

The invariance of the Euler equations and their correspondence with the NSEs also touches recent work on Onsager's conjecture \citep{Onsager1949,Eyink1995,Eyink2024,EyinkPeng2025}. \citet{DeLS2009,DeLS2010} have shown that the incompressible $3D$ Euler equations possess `wild solutions' whose kinetic energy $e(t)$ does not have to be decreasing in time and may have compact support, thus suggesting a strong degree of irregularity. These solutions, found by the method of convex integration, imply that the fluid could suddenly burst into rapid motion, and then return to rest after a finite period of time. An equivalent result has been proved for the NSEs by \citet{BV2019,BV2021}. These solutions belong to the space $W^{1,1}$ for all time, which means that both $\|\bu\|_{1}$ and $\|\nabla\bu\|_{1}$ remain finite. However, $\|\nabla\bu\|_{2}$ is not integrable in time so they do not satisfy a Leray-Hopf-type energy inequality. The suggestion that wild solutions may be the root cause of NS-turbulence is both speculative and controversial but it is nevertheless an interesting and important open question. The magazine article by \citet{EMS2025} covers a wide range of opinion by several authors over the issue of whether the Buckmaster-Vicol result is physically relevant to hydrodynamics. What is clear is that this issue has yet to be settled. A further argument is that the flaw lies in the Navier–Stokes equations themselves. The suggestion that thermal noise could make the deterministic version of the incompressible NSEs inadequate to describe the dissipation range of turbulence was first raised by \citet{Ruelle1979,Ruelle1995}. This is precisely the region covered by the PaV-scale.  More recently, \citet{Bandak2022,Bandak2024} have gone further by arguing that in molecular fluids Eulerian spontaneous stochasticity \citep{Gawedzki2008,TBM2020} is triggered by thermal noise in $3D$ Navier-Stokes turbulence at high Reynolds numbers and that this directly alters the turbulent dissipation range below the Kolmogorov scale. \citet{Bandak2024} conclude that the NSEs should be replaced by a Landau-Lifshitz-type set of equations which is comprised of the incompressible NSEs plus an additive fluctuating stress term proportional to $\mbox{div}\,\bxi$, where $\bxi$ is modelled as a mean-zero Gaussian random field with a given ensemble average. The range of scales over which it is conjectured that these thermal effects take place not only includes the dissipation range but also reaches into the inertial range, which illustrates why it is important that this issue is resolved. Moreover, if this argument turns out to be correct then it will change the direction of analysis in fluid dynamics. The search for a regularity proof of the $3D$ deterministic NSEs, although unsuccessful, has been one of \textit{the} major issues in rigorous fluid dynamics in the past generation \citep{Leray1934,FGT1981,RRS2016,JDG2019}. In contrast, the Landau–Lifschitz equations, having been designed as mesoscopic field equations with a high-wavenumber cutoff, have been proved to have strong, pathwise-unique solutions \citep{FF2008,Eyink2024}. 

%%%%%%%%%%%%%%%%%%

\section{The multifractal model of turbulence}\label{sect2:MFM}

The MFM is well-established in the literature, so we provide only the briefest of summaries \citep{Frisch1995,BJPV1998,BMV2008,Eyink2008,BB2009,BT2023}. The $p$th-order longitudinal velocity structure function $\Sp$ at a point $\bx$ in an HIT flow 
is defined as 
\bel{MFM1}
\Sp(r) = \left<\left\{\left[\bu(\bx+\br) - \bu(\bx)\right]\bcdot\hat\br\right\}^{p}\right>_{stat.av.}\,,
\ee 
where $r$ is the radius of the sphere centred at $\bx$. While the details of the statistical average are of little concern, the scaling properties of $\Sp$ in the infinite Reynolds number limit are crucial as they revolve around the invariance property of the incompressible Euler equations. %As explained in \citet{PF1985,BJPV1998,Frisch1995,BB2009}, 
The K41 formalism assumes that $S_{p}$ depends only on $r$ and the energy dissipation rate $\varepsilon = \nu L^{-3}\int_{V}|\nabla\bu|^{2}dV$. Since the product $\varepsilon r$ has the same units as $u^3$, the only scaling consistent with physical units is $S_{p} \sim (\varepsilon r)^{p/3}$. According to \citet{Frisch1995}, the symbol $\sim$ means `scales like'. This agrees with the exact result $S_{3} = - \fourfifths\varepsilon r$, which is known as the four-fifths law. While the result is elegant and satisfying, experimental and numerical data indicate that $S_p$ scales like $S_p\sim r^{\zeta_p}$, where the exponents $\zeta_p$ deviate from linear in $p$ by lying on a concave curve that, for $p\geqslant 3$, stays below the line $\sthird p$, a deviation which is attributed to intermittency \citep{Frisch1995}. To explain this, \citet{PF1985} adapted the K41 formalism by relaxing the requirement that, for small $r$, the velocity increment $\vert \bu(\bx+\br)-\bu(\bx)\vert$  scales as $r^h$ with $h=\sthird$. Instead, they allowed the scaling exponent $h$ to fluctuate.  They achieved this by writing down the probability of observing a given value of $h$ at the scale $r$ in the form
\bel{MFM2}
P_{r}(h) \sim r^{C(h)}\,,
\ee
where each value of $h$ belongs to a given fractal set of dimension $D(h)$. The co-dimension $C(h)= 3 - D(h)$ plays the role of the multifractal spectrum in the large deviation theory version expounded by \citet{Eyink2008}. Interpreting \eqref{MFM1} as an average over the probability $P_r(h)$, and by applying a steepest descent argument, one finds the following relation for the exponent $\zeta_p$
\bel{MFM3}
\zeta_{p} = \lim_{r\to 0}\left(\frac{\ln S_{p}}{\ln (\varepsilon r)}\right)\,,
\qquad\mbox{where}\qquad\zeta_{p} = \inf_{h}\left[hp + C(h)\right]\,.
\ee
Thus, $\zeta_p$ and $C(h)$ are connected through a Legendre transform. Note that when $D(h)=3$, we recover the standard K41 result that $\Sp \sim (\varepsilon r)^{hp}$. The four-fifths law requires $\zeta_{3} = 1$, thus leading to inequalities for $D(h)$ and $C(h)$
\bel{MFM4}
C(h) \geq 1 - 3h \quad\Rightarrow\quad D(h) \leq 3h + 2\,.
\ee
An alternative definition of multifractality is based on the fluctuations of energy dissipation \citep{MS1991,AFLV1992,Frisch1995}. It assumes the existence of a continuous range of scaling exponents $a$ such that the local average of the energy dissipation rate over a ball of radius $r$ scales like $r^a$ over a fractal set of dimension $F(a)$.  Kolmogorov's refined similarity hypothesis allows a connection to be established between the two definitions of multifractality; in particular, $h=a/3$ and $D(h)=F(a)$ \citep{AFLV1992,Frisch1995}. Finally, it is important to note that the MFM describes the scaling properties of the velocity field and the local energy dissipation but does not provide any information on the geometrical structure of the flow. To see how the multifractal formalism fits with the NSEs requires an examination of the $L^{2m}$-norms of the velocity gradient, which we do in the next section. 

%%%%%%%%%%%%%%%%%%%%%%%%%%%%%%%%%%%%%%%%%%%%%%%%%%%%%%

\section{The $L^{2m}$-norms of the Navier-Stokes velocity gradient}\label{sect3:NSEs}

\begin{figure}
\centering
\includegraphics[width=0.6\linewidth]{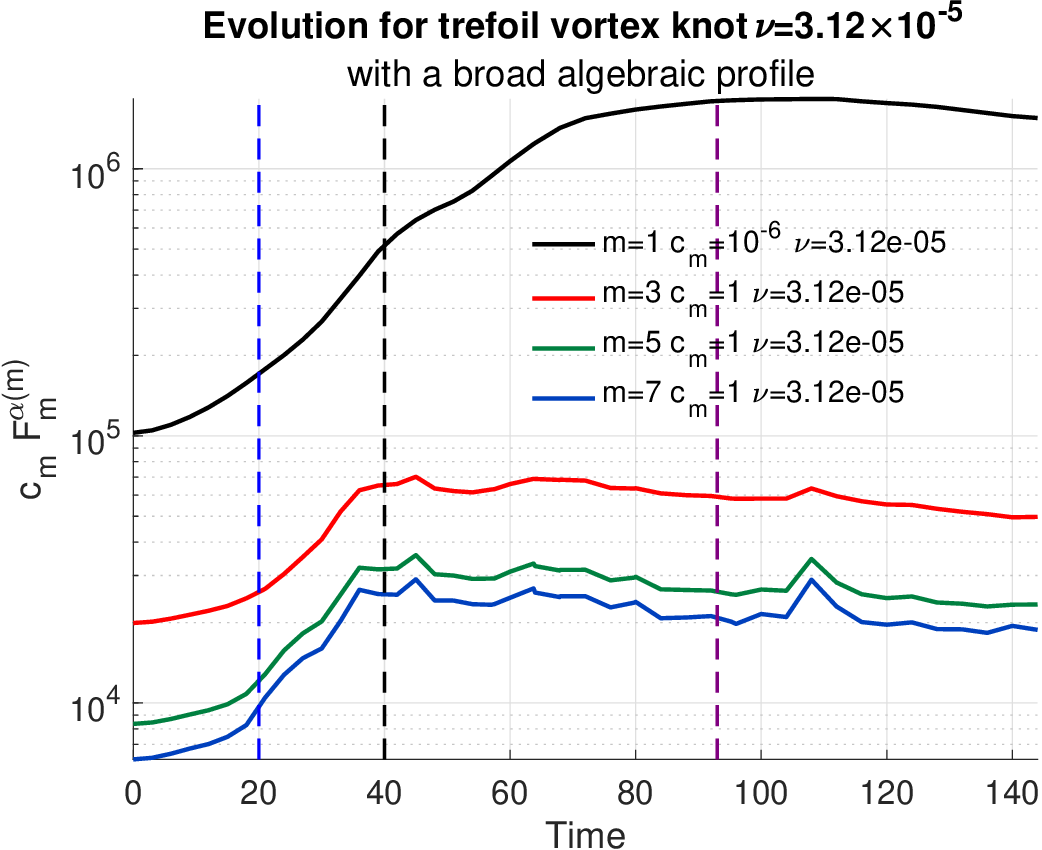}
\caption{From the evolution of a trefoil vortex knot for $\nu = 3.12\times 10^{-5}$, evidence appears of intermittent events at $t\approx 110$ for $m\geq 5$\,;  courtesy of R. M. \citet{Kerr2025}. The vertical axis is $F_{m}^{\alpha_{m}}$ defined in (\ref{nse4}) for $d=3$ which represent scaled $L^{2m}$-norms of the velocity gradient. }\label{RMK}
\end{figure}

The behaviour of the pointwise in time energy dissipation rate $\varepsilon = \nu L^{-3}\I|\nabla\bu|^{2}dV$ has been one of the primary goals in turbulence research \citep{Verma2019}.  Based on the ideas of \citet{Leray1934} and \citet{Hopf1951}, the energy inequality in a $3D$ periodic domain is
\begin{equation*}
\shalf \I \left(|\bu(\bx,\,t)|^{2} - |\bu(\bx,\,0)|^{2}\right) dV + \nu \int_{0}^{t}\I|\nabla\bu(\bx,\,\tau)|^{2}dV\,d\tau \leq 
\int_{0}^{t}\I \bu\cdot\bdf\,dVd\tau\,.
\end{equation*} 
Note that this is an inequality, not an equality, and is valid for weak solutions of the $3D$ NSEs\,: see also \citet{CKN1982} for a wider definition. Behind it lies a sophisticated theory of weak convergence properties that can be found in standard textbooks such as \citet{RRS2016}. \citet{DF2002} showed how to translate estimates based on the forcing $\bdf$ (whose $L^{2}$-norm feeds into the Grashof number $Gr$) into estimates based on the Reynolds number $Re = U_{0}L/\nu$ which is itself based on a space-time-averaged velocity field $U_{0}$, and the time-average $\left<\cdot\right>_{T}$
\bel{nse2}
U_{0}^{2} = L^{-3}\left<\|\bu\|_{2}^{2}\right>_{T}\,,\quad\mbox{with}\quad \left<\cdot\right>_{T}=\frac{1}{T}\int_{0}^{T}\cdot\,dt\,.
\ee
The Reynolds number $Re = U_{0}L/\nu$ has a solid basis because $U_{0}$ is bounded for all time. The time averaged energy dissipation rate for all smooth Navier-Stokes initial conditions and square integrable forcing is \citep{DF2002}
\bel{nse3}
\left<\varepsilon\right>_{T} = \nu L^{-3}\left<\I |\nabla\bu|^{2}dV\right>_{T} \leq L^{-4}\nu^{3}Re^{3}\,.
\ee
Inequality (\ref{nse3}) leads to the standard inverse Kolmogorov length estimate $L\lambda_{k}^{-1} \leq Re^{3/4}$, consistent with K41. However, it does not take account of strong intermittent events that are washed over by the spatial $L^2$-average in \eqref{nse3}. To address this issue we require higher $L^{2m}$-norms of the velocity gradient tensor. 
\bel{zoom}
\|\nabla\bu\|_{2m}\equiv \left(\I |\nabla\bu|^{2m}\,dV\right)^{1/2m}\,,\qquad 1 \leq m \leq\infty\,.
\ee
For values of $m$ near unity the integral is insensitive to strong intermittent events\,: the near $L^2$ spatial average washes over the most intense regions of $\nabla\bu$ with relatively little effect -- see Fig. \ref{RMK}. However, as the value of $m$ is raised, regions of greater intensity begin to dominate, and eventually at $m=\infty$ the most intense point of all completely dominates the integral.
\begin{equation*}
\begin{array}{ll}
m=1\,; & \mbox{\footnotesize corresponds to an r.m.s. average}\\
m~\mbox{\footnotesize finite~but~large}\,; & \mbox{\footnotesize corresponds to the more intense~structures}\\
m=\infty\,; & \mbox{\footnotesize corresponds to the~most intense point}
\end{array}
\end{equation*}
Thus, the sliding scale of $m = 1\to \infty$ acts as a zoom lens of a telescope. To find time-averages of $\|\nabla\bu\|_{2m}$ for $m\geq 1$ first requires a result of \citet{FGT1981}, which deserves to be better known, and which is based around $n$-th order weak derivatives within the volume integrals $H_{n}=\I |\nabla^{n}\bu|^{2}dV$. The \citet{FGT1981} result is that Leray's weak solutions satisfy 
\bel{FGT1}
\left<H_{n}^{\frac{1}{2n-1}}\right>_{T} \leq c\,L^{-1}\nu^{\frac{2}{2n-1}}Re^{3} + O\left(T^{-1}\right)\,,
\ee
in which the energy inequality is simply the case $n=1$. Using the \citet{DF2002} method, the estimate has been translated into the Reynolds number $Re$. Interpolation inequalities, together with \eqref{FGT1}, can then easily be used to find time-averages of certain powers of $\|\nabla^{n}\bu\|_{2m}$. Generalizing to other spatial dimensions $d=1,2,3$, these results can be rolled into a single formula \citep{JDG2019,JDG2020,JDG2023}. While $n\geq 3$ spatial derivatives in \eqref{FGT1} are necessary for the proof, once this has been achieved, all we need are estimates of time-averages of powers of the single derivative $\|\nabla\bu\|_{2m}$ written as two simple dimensionless formulas
\bel{nse4}
F_{m,d} = \nu^{-1}L^{1/\alpha_{m,d}}\|\nabla\bu\|_{2m}\,,\qquad\qquad \alpha_{m,d} = \frac{2m}{4m-d}\,,
\ee
where all the results rolled together can be summarized in the formula
\bel{nse5}
\left<F_{m,d}^{(4-d)\alpha_{m,d}}\right>_{T} \leq c_{m}Re^{3}\,.
\ee
This, then, is the dimensionless generalization of \eqref{nse3} to values $m \geq 1$ and $d$ dimensions. Note that when $m=1$, the factor of $4-d$ cancels and we recover the result on the energy dissipation rate in (\ref{zoom}). This result is the key not only to the correspondence between the MFM and the NSEs but also the PaV-scale. \citet{DGGKPV2014} studied the evolution of $\|\bom\|_{2m}$ numerically for 4 different data-sets. for values $m=1,\ldots,9$. Fig. \ref{RMK} is a plot of $F_{m,3}^{\alpha_{m}}$ from \citet{Kerr2025} (courtesy of R. M. Kerr) similar to its equivalent in \citet{DGGKPV2014}. The subscript $d=3$ has been suppressed. The bound in \eqref{nse5} is a rigorous Navier--Stokes result. In the next section, we will show how it can be used in connection with the phenomenological predictions of the MFM to reconcile the two approaches.

%%%%%%%%%%%%%%%%%%%%%%%%%
%%%%%%%%%%%%%%%%%%%%%%%%

\section{Two ways of reconciling the MFM and the NSEs}

%%%%%%%%%%%%%%%%%%%%%%%%%%%%%%%%

\subsection{The PaV-scale as a mediator between the MFM and the NSEs}\label{sect4:DG}
 
Having seen the properties of the Euler invariant scaling transformation in \S\ref{intro}, it is now time to apply this to the $L^{2m}$-norms of the velocity gradient defined in (\ref{zoom}).  To find a way round the problem of how to match results from the MFM with those from the NSEs, we adapt the method outlined in \citet{DG2022} where only the wavenumber $k_{pav}$ out of all possible wavenumbers was chosen, with a subsequent integration over $h$. Instead, by using the scaling transformation (\ref{rescal}) on the energy dissipation rate, we keep all wavenumbers but select the PaV-scale $k_{h,pav}=\eta_{h,pav}^{-1}$ as the only one with $h$-dependence, where the probability of observing a given exponent $h$ at the scale $\eta$ is $P_{\eta}(h)$. We adapt (\ref{MFM2}) to make it non-dimensional so that 
\bel{MFM2extra}
P_{\eta}(h) \sim \left[L^{-1}\eta\right]^{C(h)} \,,
\ee
where each value of $h$ belongs to a given fractal set of dimension $D(h)= 3- C(h)$. Our strategy is to calculate the MFM equivalent of \eqref{nse5} and compare the $\Rey$-scaling in the two formulations. This requires the application of the rescaling in (\ref{rescal}) and an average over $h$ to obtain
\bel{MFR1}
\frac{\int_{V} |\nabla\bu|^{2m}\,dV}{\int_{V}dV} 
= U_{0}^{2m}\left[\int_{h_{min}}^{h_{max}} P_\eta(h)\eta^{-2m}(L\eta^{-1})^{-2mh}\,dh\right]\frac{\int_{V'} |\nabla'\bu'|^{2m}\,dV'}{\int_{V'}dV'}\,,
\ee
in which case the $F_{m,3}$ in (\ref{nse4}) for $d=3$ becomes
\beq{MFR2}
F_{m,3} &=& \nu^{-1}L^{1/\alpha_{m,3}} \|\nabla\bu\|_{2m}\nonumber\\
&\sim& \nu^{-1}U_{0}L\left[ \int_{h_{min}}^{h_{max}} \left(L^{-1}\eta\right)^{C(h)}(L\eta^{-1})^{2m(1-h)}dh\right]^{1/2m}\left(\frac{\int_{V'} |\nabla'\bu'|^{2m}\,dV'}{\int_{V'}dV'}\right)^{1/2m}\,.
\eeq
Now we use the PaV-scale $L\eta_{h,pav}^{-1}=Re^{1/(1+h)}$ to obtain
\bel{MFR3}
F_{m,3} \sim Re\left[\int_{h_{min}}^{h_{max}} Re^{\frac{2m(1-h) - C(h)}{1+h}}dh\right]^{1/2m}
\left(\frac{\int_{V'} |\nabla'\bu'|^{2m}\,dV'}{\int_{V'}dV'}\right)^{1/2m}\,.
\ee
In the primed variables in (\ref{MFR3}), the equivalent Reynolds number is unity when $k_{pav} = \eta_{h,pav}^{-1}$ so this term is not a function of $Re$. Therefore, in the limit $Re\to\infty$, by approximating the integral over $h$ by a steepest descent argument and matching the exponents of $Re$ on the LHS with $Re^{3}$ on the right from (\ref{nse5}), we must have 
\bel{MFR4}
\alpha_{m,3}\left[1 + \max_{h}\left(\frac{2m(1-h) - C(h)}{2m(1+h)}\right)\right] \leq 3\,,
\ee
in which case
\bel{MFR5}
\min_{h}C(h) \geq 4m\left[1-3(1+h)\right] +  9(1+h)\,.
\ee
The first term on the RHS of \eqref{MFR5} must be negative to avoid $C\to\infty$ as $m\to\infty$, thus restricting $h$ to the range $h \geq - \twothirds$. Under this restriction on $h$ we need to maximize the lower bound on $C(h)$ with respect to $m$, which occurs when $m=1$. Therefore, uniform in $m$, we find that
\bel{MFR6}
C(h) \geq 1- 3h\qquad\mbox{and}\qquad h \geq - \twothirds\,.
\ee
This is consistent with the results of the four-fifths law in \S\ref{sect2:MFM} and the range of $h$ in (\ref{D2}).
\par\smallskip%\noindent
The formula in $d$ dimensions in (\ref{nse5}) incorporates a factor $4-d$ in the exponent of $F_{m,d}$ with $\alpha_{m,d} = 2m/(4m-d)$.  Therefore (\ref{MFR5}) is replaced by 
\bel{MFR7}
C(h) \geq 4m\left[1-\frac{3(1+h)}{4-d}\right] +  \frac{3d(1+h)}{4-d} \,,
\ee
in which case we require $h \geq \sthird (1-d)$ to prevent $C\to\infty$ as $m\to\infty$. Again, we must have $C(h) \geq 1- 3h$.

%%%%%%%%%%%%%%%%%%%%%%%
\subsection{A multifractal interpretation and the re-emergence of the PaV-scale}\label{sect5:tele}

We have now reached the point where we need to find a correspondence between the NSEs and the MFM. This only makes physical sense if a Navier-Stokes flow has become fully developed at large enough $T$ in the same way that \citet{EP1988} showed that the negative velocity derivative skewness must have reached a certain critical level.
\par\smallskip%\noindent
As mentioned in \S~\ref{sect2:MFM}, the multifractal formalism based on dissipation postulates the existence of fractal sets in each of which the local energy dissipation rate is scale invariant. This fact suggests a multifractal interpretation of the results from NS-analysis. Consider $\alpha_{m,d}$, defined in (\ref{nse4}) as $\alpha_{m,d} = 2m/(4m-d)$. If we simply divide through by $m$ we obtain a modified version of $\alpha$, namely $\alpha_{1,\Dim}$
\bel{nse6a}
\alpha_{m,d} = \frac{2m}{4m-d}= \frac{2}{4-\Dim} = \alpha_{1,\Dim}\,,\qquad\mbox{where}\qquad \Dim = d/m\,,
\ee
where the domain dimension $d$ has been replaced by $\Dim$. This means that working in the space $L^{2m}(\mathcal{V})$ in $d$ dimensions can be mimicked by working in the space $L^{2}(\Fm)$ in $\Dim$ dimensions. \textit{We can therefore think of $\Fm$ as a fractal set with a range of dimensions $\Dim$.} Moreover, we can rewrite the exponent $(4-d)\alpha_{m,d}$ in (\ref{nse5}) with $d$ replaced by $\Dim$ 
\bel{nse6b}
(4-\Dim)\alpha_{1,\Dim} = 2\,. 
\ee
This suggests that instead of writing down $\varepsilon$ in $L^2$ for $d=3$ as in (\ref{nse1}) we define a set of dissipation rates 
$\{\varepsilon_{m}\}$
\bel{nse7}
\varepsilon_{m} = \nu L^{-\Dim}\left<\int_{\mathbb{F}_{m}}|\nabla\bu|^{2}dV\right>_{T}\,,\quad \Dim = 3/m
\ee
for each value of $m$, on the fractal set $\Fm$. This latter may be regarded as the set in which the energy dissipation rate averaged over a ball of radius $r$ scales like $r^{a_m}$, although the dependence of the scaling exponent $a_m$ on $m$ remains unknown. Moreover, 
to obtain the full dissipation rate requires a measure over which to sum $m$, but so far this has eluded us. Noting that $\mathbb{F}_{m} \subset \mathbb{T}^{3}$, (\ref{nse7}) can be re-written as
\beq{nse8}
\varepsilon_{m} &\leq& \nu L^{3-\Dim}\left<L^{-3} \int_{\mathbb{T}^{3}}|\nabla\bu|^{2}dV\right>_{T} 
\leq  c\,L^{-1-\Dim}\nu^{3}Re^{3}\,.
\eeq
\citet{AFLV1992} wrote down the formula (\ref{nse8}) but with an arbitrary fractal dimension $D$ whereas, for each $m$, here we have a fractal set $\Fm$ with a dimension $\Dim$ that is directly associated with the NSEs.  Furthermore, $\varepsilon_{m}\nu^{-3}$ can be expressed in terms of a set of inverse lengths $\eta_{m}^{-1}$ defined such that 
\bel{nse9}
\varepsilon_{m}\nu^{-3} := \eta_{m}^{-1-\Dim}\qquad\Rightarrow\qquad 
L\eta_{m}^{-1} \leq c\,Re^{\frac{3}{1+\Dim}}\,.
\ee
Note that when $\mathfrak{D}_{1}=3$ we recover the Kolmogorov estimate $Re^{3/4}$ with a continuum of values up to $Re^3$ for $\mathfrak{D}_{\infty}=0$. 
\par\smallskip%\noindent
If $\Dim$ is interpreted as the dimension of the set $\Fm$ in which the local energy dissipation rate scales as $r^{a_m}$, it now seems natural to equate $\Dim$ in (\ref{nse7}) to the fractal dimension  $F(a_m)$ and therefore to $D(h)$. Inequality (\ref{MFM4}) leads to a simple inequality relating $m$ and $h$
\bel{D1a}
\frac{3}{m} \leq 2 + 3h\,.
\ee
We find $h \geq h_{min}$ with $h_{min} = m^{-1} - \twothirds$. The two limits $m\to 1$ and $m\to \infty$ give
\bel{D2}
-\twothirds \leq h_{min} \leq \sthird\,.
\ee
Thus $h$ is bounded away from $h=-1$. Using the inequality relation between $h$ and $m$ we can bound the exponent $3/(1+\Dim)$ in (\ref{nse9}) by
\bel{D3}
\frac{1}{1+h} \leq \frac{3}{1+\Dim}\,.
\ee
The PaV inverse scale $\eta_{h,pav}^{-1}$ thus emerges as the minimizer of the RHS for all $m$. The left hand side of (\ref{D3}) is derived directly from the MFM, while the right hand side comes from the NSEs. In (\ref{D2}), $h_{min} \geq -\twothirds$ bounds $h$ away from the dangerous value of $h=-1$, where a singularity can occur. The limit $h_{min} = -\twothirds$ corresponds to $\Dim=0$.

%%%%%%%%%%%%%%%%%%%%%%%

\begin{figure}
\setlength{\unitlength}{10mm}
\bc
\begin{picture}(6,4)(0,0)
\put(0,4.9){\makebox(0,0)[b]{Euler~Equations}}
\put(1.2,5){\vector(1,0){4}}\put(3,5){\vector(0,-1){1.5}}
\put(6.8,4.8){\vector(1,-1){1.5}}
\put(8,3){\vector(-1,1){1.5}}
\put(3.7,3.2){\vector(1,1){1.5}}
\put(5.4,4.5){\vector(-1,-1){1.5}}
\put(5.9,4.9){\makebox(0,0)[b]{NSEs}}
\put(3,2.8){\makebox(0,0)[b]{PaV}}
\put(9,2.8){\makebox(0,0)[b]{MFM}}
\end{picture}
\ec
\par\vspace{-25mm}
\caption{\footnotesize Pictorial representation of how the the PaV-scale appears as a mediator between the Euler and NSEs. Following the arrows\,: (i) Clockwise\,: application of the PaV-scale to the NSEs implies results consistent with the MFM; (ii) Anti-clockwise\,: the MFM and NSEs together imply the PaV-scale.}
\end{figure}

%%%%%%%%%%%
\section{Conclusion\,: Is our understanding of the dissipation range correct?}\label{Con}

The PaV length scale $\eta_{h,pav}$ is that value of $\eta$ in the invariant transformation of the Euler equations which, when applied to the NSEs, gives unit Reynolds number, thus implying that the inertial and dissipation terms balance. Figure 2 shows the inter-relation between the Euler equations, the NSEs, the MFM and the central intermediary role of $\eta_{h,pav}$. One of the difficulties of making a comparison between the MFM, applicable to statistically steady HIT, and the dynamic NSEs is that the weak solution methods devised by \citet{Leray1934} do not readily recognize a distinction between the inertial and dissipation ranges. Using the PaV scale $\eta_{h,pav}$ solves this difficulty by choosing $k=k_{h,pav}$ as the only value in the spectrum that is allowed to be $h$-dependent. In effect, this process auto-selects the dissipation range only. Taking the probability of observing a given scaling exponent $h$ at the scale $\eta$ to be $P_{\eta}(h) \sim \eta^{C(h)}$, where each value of $h$ belongs to a given fractal set of dimension $D(h)$, it is shown in \S\ref{sect4:DG} that a consequence of Leray's weak solution estimates leads to the inequality $C(h) \geq 1-3h$, which is consistent with the four-fifths law. It is also shown in \S\ref{sect3:NSEs} that the parameter $m$ in the $L^{2m}$-norms of the velocity gradient acts as a zoom lens focus control which is related to $h$.
\par\smallskip%\noindent
The standard picture from large-scale computational fluid dynamics is of the formation of vortical sheets, their roll-up into tubes and the breakdown of these into low-dimensional fractal segments. These sub-Kolmogorov vortical structures are usually identified with the dissipation range, in this case defined by $Lk_{h,pav} = Re^{1/(1+h)}$ with $-\twothirds \leq h \leq \sthird$. In fact, the $-\twothirds$ lower bound on $h$ corresponds to a $Re^3$ upper bound on the inverse length scale which is already close to, or deeper than, molecular scales even for modest values of $Re$. The inter-relation between the NSEs, the MFM and the central role of $\eta_{h,pav}$ identified in this paper is delicately balanced around the invariance of the Euler equations, a balance which could easily be overturned by other processes. If the suggestion by \citet{Bandak2022,Bandak2024} is correct that the thermal noise generates spontaneous stochasticity, thereby overwhelming the dissipation range, then this points to a rethink of how high-$Re$ small-scale turbulence ought to be considered. In turn it would also suggest a necessary revision of the standard computational fluid dynamics picture of low-dimensional structures associated with the dissipation range.
\par\smallskip\noindent
\textbf{Declaration of Interests\,:} The authors report no conflicts of interest.
\par\smallskip\noindent
\textbf{Acknowledgements\,:} Thanks are due to R. M. Kerr for Fig. \ref{RMK}. The authors would also like to thank the Isaac Newton Institute for Mathematical Sciences, Cambridge, for support and hospitality during the programme ``Anti-diffusive dynamics\,: from sub-cellular to astrophysical scales'' where work on this paper was undertaken. This work was supported by EPSRC grant no EP/R014604/1.
%%%%%%%%%%%%%%%%%%%%%%%%
\par\vspace{-3mm}
%%%%%%%%%%%%%%%%%%%
\bibliographystyle{jfm}
%\bibliography{jfm2esam}

\end{document}